\documentclass[12pt]{iopart}

\usepackage{graphicx}
\usepackage{dcolumn}
\usepackage{bm}
\usepackage{graphicx,rotating}
\usepackage{epsfig}

\begin{document}
 
\title{
Electronic structure of 4d transition-metal monatomic wires
}

\author{A Delin$^{1,2}$ and E Tosatti$^{2,3,4}$}
\address{$^1$Materialvetenskap, Brinellv\"agen 23, KTH, SE-10044 Stockholm, Sweden}
\address{$^2$Abdus Salam International Center for Theoretical Physics (ICTP), 
Strada Costiera 11, 34100 Trieste, Italy}
\address{$^3$International School for Advanced Studies (SISSA), via Beirut 2--4, 34014 Trieste, Italy}
\address{$^4$INFM DEMOCRITOS National Simulation Center, via Beirut 2--4, 34014 Trieste, Italy }

\ead{anna.delin@mse.kth.se}

\date{\today}
\begin{abstract}
Monatomic nanowires of the nonmagnetic transition metals Ru, Rh, and Pd
have been studied theoretically, using first-principles computational
techniques, in order to investigate the possible onset of magnetism
in these nanosystems.
Our fully relativistic spin-polarized all-electron density functional
calculations reveal the onset of Hund's rule magnetism in nanowires of all three metals,
with mean-field moments of 1.1, 0.3, and 0.7 $\mu_B$, respectively, at the equilibrium
bond length. An analysis of the band structures indicates that the 
nanocontact superparamagnetic
state suggested by our calculations should affect the ballistic conductance 
between tips made of Ru, Rh or Pd, leading to possible temperature and 
magnetic field dependent conductance.
\end{abstract}

\maketitle

\section{Introduction}
Reducing the dimensionality and size of a metallic object 
eventually leads to quantum confinement of the electrons in 
one or more dimensions.
Examples of such systems are metallic nanowires, 
where the electrons are confined in two dimensions, but unconfined
in the third dimension, along the wire.
The ultimately smallest metallic wire consists 
of just a single metallic chain of atoms.
Experimentally, long segments of such nanowires have been realized,
in particular of Au\,\cite{kondo2000_helical,rodrigues2000}. 
Production of shorter segments were recently 
reported for several other metals including the $4d$ transition 
metals Ru, Rh\,\cite{itakura2000}, and Pd\,\cite{ugarte}. 
The quantum confinement of the electrons in the wires results in
intriguing behaviour with respect to their mechanical, electrical and chemical
properties and causes
new physical phenomena to appear, for example quantized conductance\,\cite{wees1988}
and helical geometries\,\cite{gulseren1998,kondo2000_helical,tosatti2001_tension}.
Thus, the properties of these nanosystems may be dramatically 
different from the bulk properties of the same metals. 
In particular, it is interesting to explore whether and how 
nanowires of bulk nonmagnetic metals can become magnetic,
and how other properties of these nanosystems in turn are
affected by the presence of magnetism in the nanosystem, 
especially of a genuine Hund's rule magnetic order parameter.

We recently performed a similar study of the $5d$ metals Os, Ir, and Pt, which 
revealed intriguing magnetic properties of nanowire systems.
In the present paper, we concentrate on the $4d$ transition metals Ru, Rh, and Pd, and 
contrast our results for these metals with our results for the corresponding $5d$ systems.
We investigate the possibility of ferromagnetism\,\cite{note1} and its effect on 
other properties, notably quantized conductance for straight monostrand nanowires
of these metals, using state-of-the-art all-electron computational
methods based on density functional theory.
We have also performed the corresponding calculations for the noble 
metal Ag, where no magnetism is expected, for comparison. 

We address here the physics of metallic
nanowires suspended between two leads, where transmission electron microscope
images on monostrand nanowires indicate straight wire geometries\,\cite{ugarte}.
Nanowires can be stabilized in this way only temporarily, as the flow
of atoms to the leads inevitably implies stretching and thinning,
which eventually breaks the nanowire\,\cite{tosatti2001_tension}.  A free, unsuspended
chain of atoms would be totally unstable against an even larger set of
deformations, for the final stable
configuration will be a cluster, approximately spherical in shape, with a surface
dominated by close-packed facets.

In our calculations, we address strictly the straight wire geometry, with equidistant atoms.
One could imagine more complicated monostrand wire geometries, for example
zigzag geometries\,\cite{sanchezportal1999} or Peierls distortions, leading to 
di- tri- or multimerization\,\cite{peierls}.
Such distorted configurations of an unsuspended monostrand wire
may represent interesting local minima or saddle points in the total energy.
When suspended between leads, however, local minima or saddle points of
the string tension are to be considered instead of those of the energy,
since they alone will correspond to long-lived, or ``magic''
nanowires\,\cite{tosatti2001_tension}. In Au, the zigzag deformations
do not survive the string tension, and the same would happen, if they
existed, in Ru, Rh, and Pd. Thus, we shall ignore zigzag distortions, 
since they are soft against tension, in the systems we address here.
Peierls di-, tri- or multimerization distortions
are critically dependent on a long wire as well as 
on a precise  Fermi surface nesting, and
would lead to insulating nanowires.
In Ru, Rh, and Pd, the reported nanocontacts are three atoms 
long at most\,\cite{ugarte}. 
Moreover, there is no unique nesting since multiple bands cross the 
Fermi level,
the precise Fermi crossings are tension-dependent, and the corresponding
incommensurate order parameters are likely suppressed by size. The experimental evidence that
nanocontacts of Ru, Rh, and Pd are consistently metallic further suggests
neglecting Peierls distortions too until evidence to the contrary.

In wires, the electrons are confined in two dimensions.
Before investigating in more detail what effect that has
on the magnetic properties of Ru, Rh, and Pd, let us
summarize shortly what is known about the magnetic properties of these 
metals when the 
electrons are confined only in one dimension, or in all three.
 
In a monolayer grown on, or sandwiched between, 
magnetically ``inert'' substrates 
such as Cu, Ag, Au, or graphite, 
the electrons are at least approximately confined in one dimension, 
opening up the possibility
for two-dimensional magnetism.
Search for two-dimensional magnetism in Ru, Rh or Pd in such systems 
has been conducted extensively both 
theoretically and experimentally.
Starting with Ru,
a monolayer of this metal has been observed to order ferromagnetically 
when grown on graphite\,\cite{pfandzelter1995},
and when layered between graphene sheets\,\cite{suzuki2003}.
No magnetism has been observed for Ru monolayers grown on Ag or Au surfaces.
Theoretical calculations predict a Ru monolayer to be magnetic on
graphite (under certain conditions)\,\cite{chen1997,kruger1998}, 
Ag\,\cite{eriksson1991,blugel1992} and Au\,\cite{blugel1992}, 
but nonmagnetic on Cu\,\cite{garcia1999}.
What regards Rh metal, 
the only case in which two-dimensional magnetism has been observed 
is in a superlattice structure of Rh monolayers sandwiched between 
adjacent graphene sheets\,\cite{suzuki2003}.
Monolayers of Rh grown on Ag, Au, or graphite
have not shown any signs of magnetic order\,\cite{beckmann1997,chado2001,goldoni2001}.
In great contrast to the experimental results, 
Rh monolayers have been predicted to order ferromagnetically on Cu\,\cite{garcia1999},
Ag\,\cite{eriksson1991,blugel1992}, Au\,\cite{zhu1991,blugel1992}, 
and graphite\,\cite{chen1997,kruger1998}.
Finally, it has been predicted that a monolayer of Pd 
should be nonmagnetic on all substrates tested 
(Cu\,\cite{garcia1999}, Ag\,\cite{eriksson1991,redinger1995,niklasson1997}, and 
graphite\,\cite{chen1997,kruger1998}).
However, for Pd (and also Rh) films on Ag, 
calculations predict that the magnetic moment of the
film is periodically suppressed and enhanced due to quantum well effects
as a function of film thickness, 
giving rise to a finite ferromagnetic moment in certain 
films thicker than one monolayer\,\cite{niklasson1997}.

All in all, the discrepancy between 
theory and experiment regarding magnetism in 
Ru and Rh monolayers appears to be rather large at present. 
One possible explanation for this discrepancy is diffusion of
transition-metal atoms into the substrate, at least when the substrate is a noble metal.
What regards Rh on graphite, the intricacies of this system have been 
discussed in detail in reference\,\cite{goldoni2001}.

If we reduce the size of all three dimensions down to nanometer size,
we end up with clusters. 
Small Ru, Rh, and Pd clusters have been predicted to have magnetic 
ground states\,\cite{galicia1993,reddy1993,vitos2000,moseler2001}.
Experimentally, magnetism has been observed in Rh and Pd 
clusters. Counter-intuitively, large Pd clusters appear to be 
magnetic whereas small Pd clusters are not\,\cite{cox1994,sampedro2003,taniyama1997}.

Returning to magnetism in nanowires, it has been predicted that 
monatomic rows of Rh on Ag(001) are ferromagnetic,
using a semi-empirical tight-binding method\,\cite{bazhanov2000}.
Monoatomic rows of Ru, Rh, and Pd on vicinal surfaces of Ag have also been studied 
theoretically using a screened Korringa-Kohn-Rostocker 
Green function method\,\cite{bellini2001},
predicting magnetism to appear in Ru and Rh chains, but not in Pd chains.
Further, Spi{\v s}\'ak and Hafner\,\cite{spisak2003} predict ferromagnetism in Ru and Rh rows grown on
the Ag (117) vicinal surface, using the projector augmented plane-wave computational method. 
Similarly, they also find a ferromagnetic ground state for Ru rows grown on Cu (117), and for
freely hanging Ru and Rh mono-strand nanowires.

Regarding the objects we study in this paper, 
i.e., monostrand wires hanging between leads, 
we can actually get some first clues about possible magnetism
simply by starting from the atomic ground state of Ru, Rh, Pd, and Ag and 
analyzing the effect of a very weak hybridization of orbitals on adjacent atoms.
In wire form, Ru and Rh could conceivably
develop Hund's rule magnetism, due to their partially empty narrow band-width $d$ shell. 
The Pd atom has in fact a filled $4d$ shell, but $sd$ hybridization in the wire enables 
$d\rightarrow s$ electron transfer, opening up 
the possibility to spin polarize Pd $d$-holes.
Ag, on the other hand, is basically an $sp$ metal, 
with the $4d$ shell completely filled in all cases.
Interestingly, it is in principle possible that wires of metals like Ag 
(a typical system that might be thought of as a jellium) 
in themselves magnetize under certain 
conditions, since even a jellium confined in a thin cylinder in principle
magnetizes for certain radii of the cylinder\,\cite{zabala1998}.
However, the moment formation in that case is weak and confined to 
very special radii or electron densities,
and the associated energy gain is very small. 
That is of course so because exchange interactions, as 
described by Hund's first rule, are not particularly
strong in an $sp$ band metal or jellium. 
The situation is radically different for transition metals. 
Because of the partly occupied $d$ orbitals, their ability to magnetize is 
much stronger and of a fundamentally different nature compared to the jellium.

In the discussion above we have completely neglected the 
issue of fluctuations.
Thermal fluctuations in a nanowire are expected to be very large, 
and would destroy long range magnetic order in the absence of an external 
magnetic field.  
In earlier papers\,\cite{delin_5d_wires,delin_pd_wire}, we have argued in more 
detail how one might deal with fluctuations, and in what cases
one can approximate some properties of the superparamagnetic state with
those of a statically magnetized one (which is what we
calculate), and we will not repeat those arguments here.
Experimentally, evidence of 
one-dimensional superparamagnetism  with fluctuations sufficiently slow on the time 
scale of the probe has been recently reported for Co atomic chains deposited on Pt 
surface steps\,\cite{gambardella2002}.

 \begin{figure}[h]
\centerline{
\includegraphics[scale=0.6,angle=0]{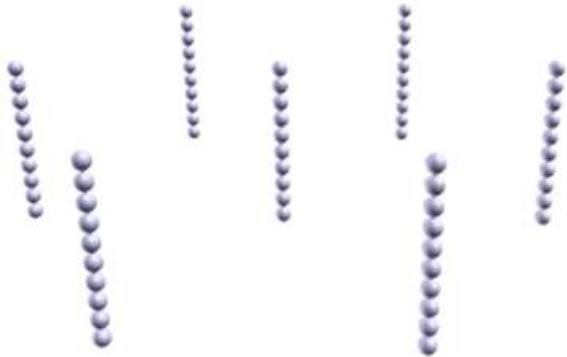}
}
 \caption
  {
Sketch\,\cite{xcrysden} of the setup. Infinitely long wires extend
along the $z$-direction and form a hexagonal mesh in the $xy$-plane.
 \label{fig:figure_1}
  }
 \end{figure}
\section{Method}
The technical aspects of the present calculations are similar to those
reported in our earlier papers on $5d$ metal nanowires\,\cite{delin_5d_wires}.
For the present density-functional-based\,\cite{dft} electronic-structure calculations
we used the all-electron full-potential linear
muffin-tin orbital method (FP-LMTO)\,\cite{wills}.
This method assumes no shape approximation of the potential or wave functions.
The calculations were performed using the generalized gradient approximation
(GGA)\,\cite{gga}.
As a test, some calculations were also performed using the local density
approximation (LDA)\,\cite{lda}, giving results very similar to the GGA ones. 
Further, some calculations were double-checked using the WIEN code\,\cite{lapw,wien97}, again with 
very similar results.

The calculations were performed with inherently three-dimensional
codes, and thus the system simulated was an infinite two-dimensional
array of infinitely long, straight wires.
Figure~\ref{fig:figure_1} shows a sketch of the setup.
A one-dimensional Brillouin zone was used, i.e 
the k-points form a single line, stretching along the 
$z$-axis of the wire. 
The Bravais lattice in the $xy$-plane was chosen hexagonal.
With this choice, the $d_{xy}$ and $d_{x^2 - y^2}$ orbitals become
automatically degenerate, as they should for a single wire. 
Furthermore, we used non-overlapping muffin-tin spheres with a 
constant radius in the calculations
of the equilibrium bond lengths $d$. 
The magnetic moments, bands structures, conductance-channel 
curves and  band widths were calculated using 
muffin-tin spheres scaling with the bond length. 
Convergence of the magnetic moment was 
ensured with respect to k-point mesh density, Fourier mesh density,
tail energies, and wire-wire vacuum distance.

We performed both scalar relativistic (SR) calculations, and calculations 
including the spin-orbit
coupling as well as the scalar-relativistic terms. 
The latter will be referred to as ``fully relativistic'' (FR) 
calculations in the following,
although we are not strictly solving the full Dirac equation, or 
making use of current density functional theory.
In the fully relativistic calculations, the spin axis was chosen to be 
aligned along the wire direction.

\begin{table}
 \caption{
 \label{tab:lattpar}
  Calculated bulk and wire equilibrium bond length $d$.
  Also shown are the wire magnetic moments per atom,
  calculated at the equilibrium bond length.
  The right-most column displays the experimental ground-state
  configuration for the free atoms.
  (FR = fully relativistic calculation; SR = scalar relativistic calculation)
 }
 \begin{indented}
 \lineup
 \item[]\begin{tabular}{cccccl}
 \br
       & wire         & bulk         &  bulk         & moment        & free                     \\
       &  $d$ ({\AA}) & $d$ ({\AA})  &  $d$ ({\AA})  & ($\mu_B$)     & atom                     \\
 metal & FR           & FR           &  exp.         & FR;SR         & configuration             \\
 \mr
    &                 &              &               &               &                   \\
 Ru & 2.27            & 2.70         & 2.71          & 1.1           &  4 ($^5 F_{5}$)          \\
 Rh & 2.31            & 2.72         & 2.69          & 0.3           &  3 ($^4 F_{9/2}$)        \\
 Pd & 2.56            & 2.78         & 2.75          & 0.7           &  0 ($^1 S_{0}$)          \\
 Ag & 2.68            & 2.93         & 2.89          &  -            &  1 ($^2 S_{1/2}$)        \\
 \br
 \end{tabular}
                                                                              
 \end{indented}
\end{table}

\section{Results}

\subsection{Bond lengths and energetics}
In a monostrand nanowire, there are only two nearest neighbours, 
and therefore we expect the
bond length minimizing the total energy to be smaller than in the bulk.
This is indeed the case, as can be seen in Table~\ref{tab:lattpar}, where calculated
bond lengths for monowires and bulk are listed, 
together with the experimental bulk values.
Our bulk GGA calculations for the equilibrium bond lengths are in very close agreement
with the experimental values. 
Our calculated bond lengths for free-standing, monostrand wires 
are close to (within a few hundredths of an {\AA}) the 
ones reported in reference\,\cite{bahn2001} (Pd and Ag), but significantly
larger than (the differences are of the order 0.1 {\AA}) the ones reported in 
reference\,\cite{spisak2003} (Ru, Rh, and Pd).
We also wish to point out here that strictly speaking, a tip-suspended 
wire will not have a quasi-stable configuration
at the bond length which minimizes the total energy, but at a slightly
larger value since it is rather the string tension than the total energy 
which should attain a local
minimum\,\cite{tosatti2001_tension}. Nevertheless, for simplicity, in the remainder of 
this paper, the bond length which minimizes the total energy will be called the
equilibrium bond length. 

Table~\ref{tab:lattpar} also shows our calculated mean-field magnetic 
moments per atom at the equilibrium bond lengths.
Note that the scalar relativistic and fully relativistic calculations 
predict the same magnetic moments within the precision given
at the equilibrium bond length. 
Thus, the spin-orbit coupling appears to be unimportant
what regards the existence and magnitude of the magnetic 
moments in the $4d$ metals Ru, Rh, and Pd. 
This makes a strong contrast to the situation in $5d$ nanowires of Os, Ir, and Pt, 
where relativistic effects were shown to be
crucial for the correct description of the magnetic profiles\,\cite{delin_5d_wires}.
Even for the $4d$ metal nanowires, however, it turns out that the spin-orbit coupling
is by no means  unimportant, as will be further elucidated in the
analysis of the energetics, band structures, and conductance channels.
Calculated magnetic moments for straight monostrand 
wires of Ru and Rh have been reported also by 
Spi{\v s}\'ak and Hafner\,\cite{spisak2003}. 
They found 0.98 $\mu_B$ for Ru, and 0.26 $\mu_B$ for Rh, which is
in excellent agreement with our calculated spin moments.
Our result for Pd differs from that of Bahn {\it et al}\,\cite{bahn2001} who
found no magnetism in pseudopotential calculations for Pd monostrand nanowires.
It seems possible that the disagreement could arise in this very
borderline case due to the different methods
used, in which case we would tend to trust our all-electron approach better.
Experiments will have to be awaited in order to settle this question.

The right-most column in Table~\ref{tab:lattpar} lists the experimental atomic ground
state configuration, showing that the free Ru, Rh, Pd,and Ag atoms have 
spin moments of
4, 3, 0, and 1~$\mu_B$, respectively. Thus, we see that the 
predicted wire moments at the 
equilibrium bond length are much smaller than the magnetic 
moments of the free atom, except for 
Pd, where the atomic moment is zero, but the wire has a substantial magnetic moment of 
around 0.7 $\mu_B$.

\begin{table}
\caption{
\label{tab:energies}
Total energy difference between wire and bulk, and between the
nonmagnetic and ferromagnetic wire, per atom.
(SR = scalar relativistic calculation; FR = fully relativistic calculation;
NM = nonmagnetic calculation; FM = ferromagnetic calculation)
}
 \begin{indented}
 \lineup
 \item[]\begin{tabular}{cccccc}
 \br
           & $E_{\rm wire} - E_{\rm bulk}$   &  $E_{\rm wire} - E_{\rm bulk}$  &  $E_{\rm NM} - E_{\rm FM}$  &  $E_{\rm NM} - E_{\rm FM}$  \\
           &      (eV)                       &       (eV)  &      (meV)                  &     (meV)                   \\
 metal     &       SR                        &        FR  &       SR                    &      FR                     \\
 \mr
           &                                 &  &                             &                             \\
 Ru        &       5.4                       &      4.9  &    77                       &     56                      \\
 Rh        &       4.7                       &      4.1  &    10                       &     \09                      \\
 Pd        &       3.1                       &      3.1  &    25                       &     12                      \\
 Ag        &       1.8                       &      1.8  &     -                       &      -                      \\
 \br
 \end{tabular}
 \end{indented}
\end{table}
In order to analyze the relative stability of wire formation,
we calculated the energy difference between wire and bulk. 
The results are displayed in Table\,\ref{tab:energies}.
The energy difference between wire and bulk is smallest for Ag (1.8\,eV), and increases
as one goes left in the 4d series to Pd, Rh and Ru. 
Spin-orbit coupling somewhat reduces this energy 
difference with about 0.5\,eV for Ru and Rh, 
whereas it has no effect in Pd and Ag.
Scalar-relativistic energy differences between monowire and bulk 
have been reported earlier for Pd and Ag, and our numbers 
agree well with that calculation\,\cite{bahn2001}.

We also calculated the energy gain when the wire is allowed to 
spin polarize, $E_{\rm NM} - E_{\rm FM}$.
This quantity is typically a few hundredths of an eV,
and differs greatly from element to element.
Spin-orbit coupling halves $E_{\rm NM} - E_{\rm FM}$ 
in the case of Pd, whereas the effect of spin-orbit coupling on magnetism is
much smaller for Rh and Rh.
In reference\,\cite{spisak2003}, this energy difference 
was reported to be 39\,meV for Ru, and 
6\,meV for Rh, i.e., smaller compared to the ones calculated in the 
present work (56 and 9\,meV, respectively).

\subsection{Magnetic moments and band structures}
\begin{figure}[h]
\centerline{
\includegraphics[scale=0.8,angle=0]{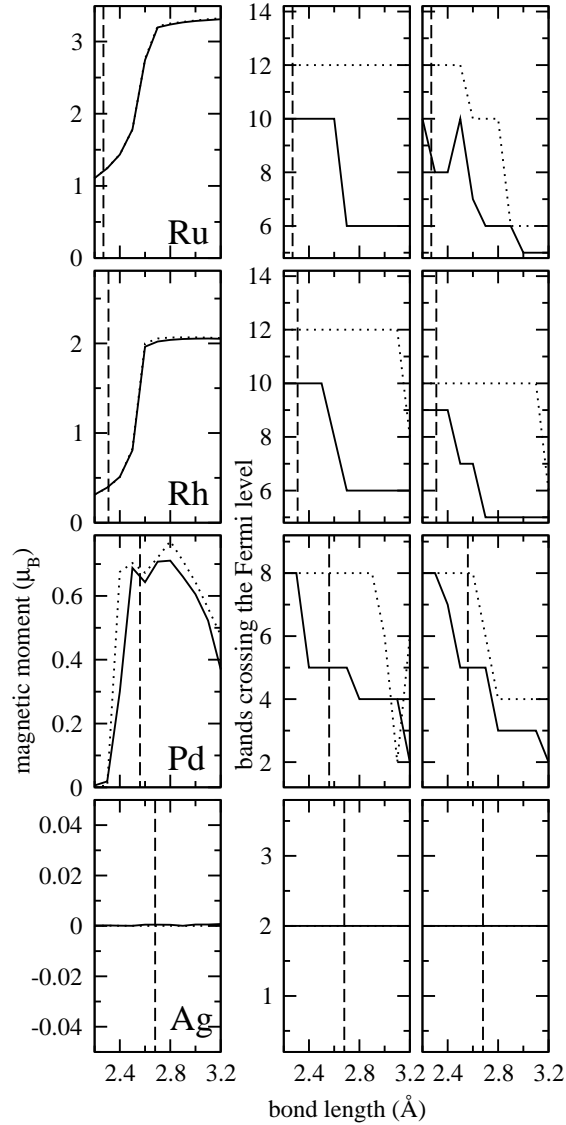}
}
 \caption
  {
Magnetic profiles and number of bands crossing the Fermi level
for the four metals studied.
The left-most column shows total magnetic moments per atom,
as a function of bond length, both with spin-orbit coupling
(FR, solid line) and without (SR, dotted line).
The middle and right-most columns show the number of
bands crossing the Fermi level
as a function of bond length for the SR and FR calculation,
respectively. The solid lines refer to ferromagnetic calculations,
and the dotted lines are for nonmagnetic calculations.
The dashed vertical lines point out the equilibrium bond lengths.
 \label{fig:figure_2}
  }
 \end{figure}

The magnetic moments per atom as a function of nanowire bond 
length are shown in the left-most 
column of figure~\ref{fig:figure_2}.
The solid lines refer to the fully relativistic calculations, 
and the dotted lines to the scalar relativistic calculations.
All the metals studied, except Ag, 
exhibit a magnetic moment for values of the 
bond lengths at equilibrium. 
Spin-orbit coupling apparently has a very limited effect on the 
magnetic profiles; for Ru and Rh, the
difference is not even visible; for Pd however there is a small 
but visible quantitative difference. 

The magnetic profiles for Ru and Rh are quite similar to each other. 
For both metals, the 
magnetic moment increases with stretching, and reaches a 
plateau value for large bond lengths.
The magnetic profile of Pd, on the other hand, is unique in that it has a maximum
for bond lengths around the equilibrium value, and then decreases down to zero for
stretched bond lengths (see also reference\,\cite{delin_pd_wire} for a more in-depth
discussion of the Pd nanowire magnetic profile). 
The Ag wire is, unsurprisingly, firmly nonmagnetic, for 
the bond lengths studied.

In order to shed some light onto the mechanisms behind the 
magnetic profiles displayed in
figure~\ref{fig:figure_2}, we will now 
analyze the electronic structure of the wires with help of 
the band structures at different bond lengths.

 \begin{figure}[h]
\centerline{
\includegraphics[scale=0.8,angle=0]{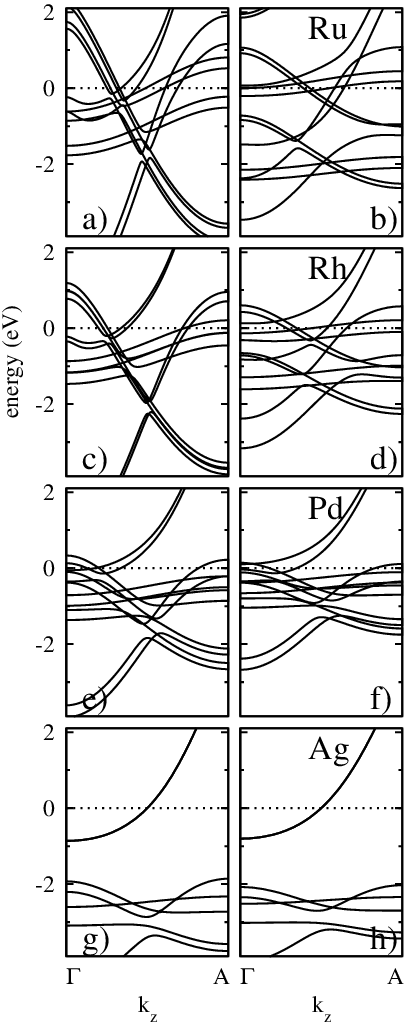}
}
 \caption
  {
FR band structures, along the wire direction,
at two different bond lengths
(the equilibrium one, and a larger of 2.9 {\AA}) for each element.
The Fermi energy is at zero.
Band doubling (present in panels a through f) indicates spin splitting due
to magnetic order.
 \label{fig:band_structure_fr}
  }
 \end{figure}
 \begin{figure}[h]
\centerline{
\includegraphics[scale=0.8,angle=0]{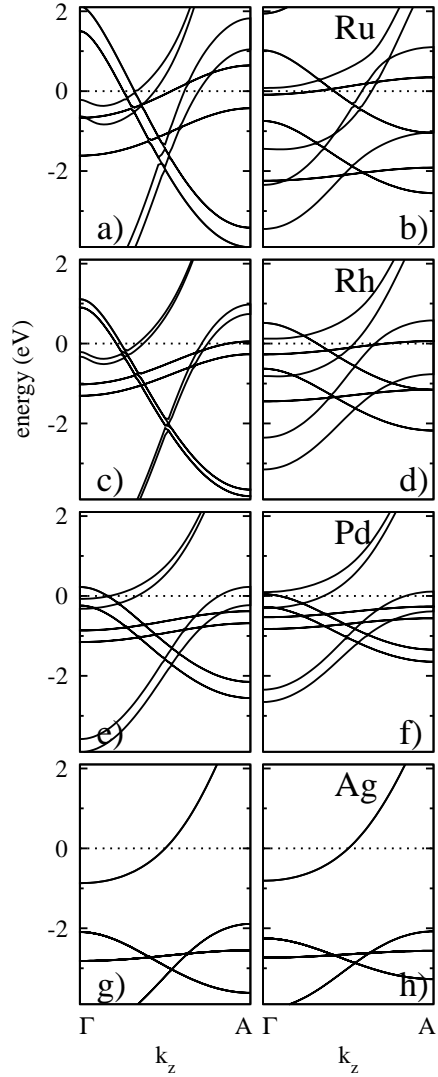}
}
 \caption
  {
SR band structures, along the wire direction,
at two different bond lengths
(the equilibrium one, and a larger of 2.9 {\AA}) for each element.
The Fermi energy is at zero.
Band doubling (present in panels a through f) indicates spin splitting due
to magnetic order.
 \label{fig:band_structure_sr}
  }
 \end{figure}
 \begin{figure}[h]
\centerline{
\includegraphics[scale=0.8,angle=0]{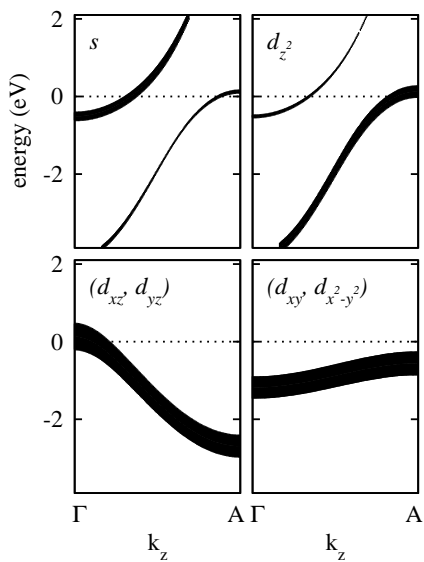}
}
 \caption
  {
Character-resolved SR band structure for
non-spin-polarized Pd, along the wire direction.
The Fermi energy is at zero.
The $d_{xz}$ and $d_{yz}$ orbitals are always degenerate,
and the same is true for the
orbital pair $d_{xy}$ and $d_{x^2-y^2}$.
 \label{fig:fatbands}
  }
 \end{figure}
Band structures for two different bond lengths, 
the equilibrium bond length, and a larger one of 2.9 {\AA}, 
roughly representing two magnetic regimes, 
are shown in figure~\ref{fig:band_structure_fr} (fully relativistic calculation) 
and figure~\ref{fig:band_structure_sr} (scalar-relativistic calculation)
for each of the four metals studied.
The bands run from the zone center, $\Gamma$, to the zone edge, A, 
in the direction of the wire.

The character of the bands close to the Fermi level is of 
critical importance for the moment formation,
and therefore we also show character-resolved bands, 
see figure~\ref{fig:fatbands}. 
We found it useful to split up the $d$ character into three contributions:
$d_z$, $(d_{xz},d_{yz})$ 
and $(d_{xy},d_{x^2-y^2})$. Thus,
figure~\ref{fig:fatbands} has four panels, displaying separately the 
$s$, $d_z$, $(d_{xz},d_{yz})$ and $(d_{xy},d_{x^2-y^2})$ characters
of the bands.
The vertical error bars, or ``thickness'', of the bands 
indicate the relative character weight. 
The data in figure~\ref{fig:fatbands} have been taken from a 
scalar-relativistic calculation for Pd.
For the other metals, the relative weights of the orbitals for 
each band are qualitatively similar to the ones shown. 
From figure~\ref{fig:fatbands}, we see that most bands in the 
vicinity of the Fermi level are of predominantly $d$ character. 
In fact, there are only two bands with some $s$ character 
crossing the Fermi level (see upper left
panel in figure~\ref{fig:fatbands}). Of these, the highest lying band 
crosses the Fermi level  closer to the zone center than the zone edge, at about
one third of the distance between the zone center and zone edge.
The second one of the two $s$-containing bands crosses the Fermi level very close to the zone edge (A).
At that point, this band has some $s$ character, but 
is in fact dominated by $d_z$ character.

At  $\Gamma$ and A, (both critical points by symmetry), 
all band dispersions are horizontal. 
This gives rise to very sharp $1/\sqrt{E}$ band edge
van Hove singularities, due to the one-dimensionality of the systems. 
If a band has mostly $d$ character at the edge, the exchange energy
gain will be rather large if the band spin-splits so that one of 
the spin-channel band edges ends up above the Fermi level, and the other one below.
We note in passing that, strictly speaking, 
in the fully relativistic calculations (figure~\ref{fig:band_structure_fr})
the spin-orbit coupling will mix the two spin channels 
so that, in general, an eigenvalue
will have both majority and minority spin character. 
However, in the present calculations, this mixing is 
so small, typically just a few percent, that it is
irrelevant for our qualitative discussion here.
Thus, if a band edge ends up sufficiently near the Fermi level, 
we may expect a magnetic moment
to develop. While apparently similar to the magnetization of 
the jellium wire\,\cite{zabala1998}, magnetism here is much 
more substantial, since here the $d$ states involve
a much stronger Hund's rule exchange.  
We now go through all four metals, starting with Ru, 
analyzing how the band edges move as a function of bond length,
and how this affects the magnetic state of the wires.

{\it Ru:}
The magnetic moment of Ru increases with the bond length.
At the equilibrium bond length, 
the rather flat $(d_{xy},d_{x^2-y^2})$ bands are split around the 
Fermi level, creating a
relatively large magnetic moment of 1.1 $\mu_B$, see
panel a in figures~\ref{fig:band_structure_fr} and \ref{fig:band_structure_sr}.
The $(d_{xz},d_{yz})$ bands are still broad at this bond length and 
in principle unpolarized.
As the bond length increases, the $(d_{xz},d_{yz})$ bands narrow down and eventually
also spin-polarize. They split around the Fermi level, and causes a large 
magnetic moment of more than 3 $\mu_B$ for bond lengths larger than 2.7 {\AA}.

{\it Rh:}
As in Ru, the magnetic moment of Rh increases with the bond length.
In fact, the magnetic profile for Rh is very similar to the one of Ru, 
just shifted in magnitude.
With one more electron than Ru, the bands of the Rh wire lie generally deeper. 
The result is that the flat $(d_{xy},d_{x^2-y^2})$ bands, 
just barely touch the Fermi level, instead of clearly crossing it, as in Ru.
The result is a smaller splitting, and consequently a smaller moment.
With increasing bond length, all the other bands also narrow down and eventually
split around the Fermi level,
just like for Ru. The result is a plateau value of the magnetic moment of
about 2 $\mu_B$ for bond lengths above 2.6 {\AA}.
Thus, in both Ru and Rh, the flat $(d_{xy},d_{x^2-y^2})$ bands drive the 
formation of the magnetic moment, and the $(d_{xz},d_{yz})$ bands enhance it. 

{\it Pd:}
In Pd, the very same flat $(d_{xy},d_{x^2-y^2})$ bands leading to Hund's
rule magnetism in Ru and Rh are entirely occupied at all bond lengths studied.
The spin polarization is instead driven by the  $s+d_{z^2}$ and
$(d_{xz},d_{yz})$ bands, which have a high dispersion, and
display one-dimensional band edges close to the Fermi level at $\Gamma$ and A.
In the magnetic regime, it happens that
these three band edges all are nearly degenerate
in energy and close to the Fermi level.
This accidental feature
of the long Pd monostrand nanowire band structure dramatically increases the density
of states, since divergent van Hove singularities are formed close to
the Fermi level. 
For the stretched wire with bond length 2.9 {\AA} 
(panel f in figures~\ref{fig:band_structure_fr} and \ref{fig:band_structure_sr}), 
the $s$-dominated band with band edge at $\Gamma$
is split around the Fermi level.
A more detailed discussion of the physics of Pd nanowires 
can be found in reference\,\cite{delin_pd_wire}.

{\it Ag:}
For Ag, basically an $sp$ metal,
all the $d$ bands are filled and there is never any $d$-magnetism.
The only band crossing the Fermi level is one of the $s$-dominated, high-dispersion, 
$s+d_{z^2}$ bands.
Calculations have shown that monostrand wires of $sp$ metals, and even a jellium cylinder,
can spin-split\,\cite{zabala1998, ayuela2002}.
For this to happen however, a band edge must be extremely close to the Fermi level.
The $s+d_{z^2}$ band in the Ag wire has a band edge at $\Gamma$ approximately 1\,eV 
below the Fermi level, which is too far down to make a spin splitting possible.

\subsection{Ballistic conductance channels}

As seen from the above discussion of the nanowire band structures,
spin-splitting of bands does alter $n$, i.e., the number of bands (or channels) 
crossing the Fermi level.  By virtue of the Landauer formula 
\begin{equation}
G = \frac{e^2}{h}\sum_i \tau_i,
\end{equation}
where $\tau_i$ is the transmission through channel $i$,
the maximum theoretical ballistic conductance has, 
in units of $\frac{1}{2}G_0 = e^2/h$, precisely the
number of bands $n$ crossing the Fermi level as its upper limit.
Thus, the conductance through the wires should be affected by 
the presence of magnetism.

The middle column of figure~\ref{fig:figure_2} (fully relativistic calculation) 
and right-most column of figure~\ref{fig:figure_2} (scalar relativistic calculation) 
illustrate how
$n$ is influenced by nanowire spin-polarization, bond length, and
spin-orbit coupling.
For Ru, Rh, and Pd in their magnetized state at the equilibrium bond length, 
$n$ is 9, 9, and 5, respectively, compared to 12, 10, and 8 
in the nonmagnetic state when spin-orbit coupling is included in the calculation. 
In the scalar-relativistic calculations, the channel count at 
the equilibrium bond length for Ru and Rh is 
close but not identical to the channel count of the 
fully relativistic calculation. For Pd
the two calculations give the same channel count at equilibrium bond length.
For stretched wires, however, the scalar and fully relativistic calculations
differ also for Pd in the number of conductance channels.

In general, spin-polarization tends to decrease the number of channels. 
Should all the channels transmit fully, large ballistic conductances of
$4.5 G_0$, $4.5 G_0$, and $2.5 G_0$ for Ru, Rh, and Pd, respectively would 
ensue, to be compared with
nonmagnetic conductances of $6 G_0$, $5 G_0$, and $4 G_0$, respectively. 

In reality however, the situation will be quite different, due to the fact that
most of the open channels have $d$ character. While the
conductance of the $s$-dominated channels is generally close to one owing to nearly
complete transmission, that of the $d$-channels is much
smaller, with a high reflection at the lead-wire junction, 
generally dependent on the detailed junction geometry. 
Our calculations reveal that the
occupation of each of the two $s$-channels is always finite for all bond lengths 
reported here, bringing an expected contribution close to $G_0$ 
to the total conductance.
The $d$-channel contribution to the conductance
is expected to be much smaller than $\frac{1}{2} G_0$ per channel.
All in all, we may thus expect monatomic nanocontacts of Ru, Rh, and Pd 
to have a conductance above $G_0$ but well 
below $4.5 G_0$, $4.5 G_0$, and $2.5 G_0$, respectively.
Since the scattering of the $d$ waves at the junctions depends highly 
on the geometry, whose details will change at every realization, 
we also expect the conductance histograms to exhibit peaks 
that could be both broad and poorly reproducible. 

Conductance histograms and traces for Ru and Rh nanocontacts
have been measured by Itakura {\it et al}\,\cite{itakura2000}.
In both these metals, they find a broad bump between $G_0$ and $2 G_0$,
which is consistent with our results.

Enomoto {\it et al}\,\cite{enomoto2002}
have published conductance histograms and traces for Pd-Ag alloys.
In their conductance histogram for pure Pd, there is no significant
structure in the region $G_0$ - $4 G_0$, whereas for Ag, there is a sharp
peak at $G_0$, as expected. 
Rodrigues {et al.}\,\cite{ugarte} have measured the 
conductance through pure Pd nanocontacts in
ultra-high vacuum, and found a peak at very low conductance, around  $0.5 G_0$.

\section{Discussion and conclusions}
Our all-electron calculations suggest that the Ru, Rh, and Pd monatomic
nanowires exhibit spontaneous Hund's rule magnetism for values of the
bond length at and around equilibrium.
The energy gain connected with the magnetic state is of the order of a few
hundredths of an eV.
This indicates that the magnetization could be stabilized against fluctuations
at cryogenic temperatures, especially with the help of an external field.
From a methodological point of view, 
the spin-orbit coupling is found to be 
important for a correct description of the energetics,
and the number of $d$-dominated conductance channels.
On the contrary, the spin moments themselves 
are rather insensitive to the spin-orbit coupling.

How might this nanomagnetism be detected experimentally? 
Merely measuring the conduction through the wire 
at one single temperature and magnetic field strength will most probably not give
conclusive information regarding the magnetic state of the atoms in the wire, since
the transmission through $d$-channels is rather poor and 
vary greatly with geometry and can hardy be regarded as quantized.
We speculate that the conductance may vary in the following qualitative way as
a function of temperature and magnetic field strength.
First, at low temperature and zero field a Kondo state between the conduction electrons
and the nanomagnet could form, implying a high ballistic conductance\,\cite{costi}.
Second, at high temperature and zero field
Kondo-like effects could affect the ballistic conductance and 
cause it to drop\,\cite{costi}.
Third, low temperature and a high magnetic field would 
take the nanowire to a magnetic, 
or in any case to a slowly fluctuating superparamagnetic regime. 
In this regime the number of conductance channels
should diminish, and so should the conductance.
At sufficiently low temperatures, the conductance should therefore be field sensitive,
and that magnetoresistance would be a clear indication of a magnetic state.
Therefore, a key experiment would be to measure ballistic conductance as a function of 
both temperature and external magnetic field. 
 
Fractional conductance peaks below $G_0$ have been observed 
experimentally, for example the $\frac{1}{2}G_0$ peak reported
by Ono for Ni\,\cite{ono1999}, and very recently by Rodrigues {\it et al} for 
Co, Pd and Pt\,\cite{ugarte}, at room temperature and zero field. 
These results are intriguing, since we expect that the $s$-channels 
alone should yield a conductance larger than that.
Impurities could be a possible explanation for these low-conductance peaks,
as Untiedt {\it et al}\,\cite{untiedt2003} very recently demonstrated
in the case of Pt. But several questions remain, such as why the $\frac{1}{2}G_0$ peak
is much larger in Co and Pd than in Pt\,\cite{ugarte}.
Therefore, it would be highly interesting to see if the half-conductance peak
exists also in conductance histograms of Ru and Rh, and how the relative size of 
that peak, if it exists, varies with impurity concentration.
We discussed in previous work\,\cite{smogunov2002}, a
possibility to obtain conductance $G_0$ from a magnetic transition metal
nanowire with a magnetization reversal occurring inside the nanowire. 
This could in principle drop to $\frac{1}{2}G_0$ in an asymmetrical situation, 
with a net prevalence of majority spins over minority spins. 
We are however at the present time not able to explain how that kind of state could be
sustained in Pd, at the experimental conditions
of zero field and room temperature. 
Finally, we also note that the conductance histogram peaks 
in reference\,\cite{ugarte} for Co, Pd, and Pt, 
centered around  $\frac{1}{2}G_0$, are rather broad, which 
suggests that they might not be caused by one single fully 
transmitting spin-polarized channel, but perhaps by
several poorly conducting channels.
In any case, more theory work will be needed to address 
the experimental data,
explicitly including such elements as tip form, 
spin structures, impurities\,\cite{untiedt2003},
strong correlations, and temperature as well
as their effects on the system's conductance.

\ack
A.D. acknowledges financial support from 
the European Commission through contract no. HPMF-CT-2000-00827,
STINT (Swedish Foundation for International Cooperation in 
Research and Higher Education),
and VR (Swedish Research Council).
Work at SISSA was also sponsored through TMR FULPROP, MIUR (COFIN and FIRB) 
and by INFM/F.
Ruben Weht is acknowledged for discussions, and for double-checking some of the 
calculations using the WIEN97 code.
J. M. Wills is acknowledged for letting us use his FP-LMTO code.
We are also grateful to D. Ugarte for sharing with us the 
results of reference\,\cite{ugarte}
prior to publication, and to C. Untiedt for several discussion.


\begin{thebibliography}{99}

\bibitem{kondo2000_helical}
Kondo K and Takayanagi K
2000 {\it Science} {\bf 289} 606

\bibitem{rodrigues2000}
Rodrigues V, Fuhrer T and Ugarte D 
2000 {\it Phys. Rev. Lett.} {\bf 85} 4124

\bibitem{itakura2000}
Itakura K, Yasuda H, Kurokawa S and Sakai A
2000 {\it J. Phys. Soc. Japan} {\bf 69} 625

\bibitem{ugarte}
Rodrigues V, Bettini J, Silva P C and Ugarte D
2003 {\it Phys. Rev. Lett.} {\bf 91} 096801

\bibitem{wees1988}
van Wees B J, van Houten H, Beenakker C W J, Williamson J G,
Kouwenhoven L P, van der Marel  D and Foxon C T
1988 {\it Phys. Rev. Lett.} {\bf 60} 848

\bibitem{gulseren1998}
G\"ulseren O, Ercolessi F and Tosatti E
1998 {\it Phys. Rev. Lett.} {\bf 80} 3775

\bibitem{tosatti2001_tension}
Tosatti E, Prestipino S, Kostlmeier S, Dal Corso A and Di Tolla F D 
2001 {\it Science} {\bf 291} 288

\bibitem{note1} In the case of Pd, we also performed 
antiferromagnetic calculations, but
found this magnetic configuration to be unstable
with respect to ferromagnetic ordering.

\bibitem{sanchezportal1999}
Sanchez-Portal D, Artacho E, Junquera J, Ordej\'on P, Garc\'{\i}a A and 
Soler J M
1999 {\it Phys. Rev. Lett.} {\bf 83} 3884

\bibitem{peierls}
Peierls R E, 1955 {\it Quantum theory of solids} 
(London: Oxford University Press)

\bibitem{pfandzelter1995}
Pfandzelter R, Steierl G and Rau C 
1995 {\it Phys. Rev. Lett.} {\bf 74} 3467

\bibitem{suzuki2003}
Suzuki M, Suzuki I S and  Walter J
2003 {\it Phys. Rev. }B {\bf 67} 094406

\bibitem{chen1997}
Chen L, Wu R, Kioussis N and Blanco R J 
1997 {\it J. Appl. Phys.} {\bf 81} 4161

\bibitem{kruger1998}
Kr\"uger P, Parlebas J C, Moraitis G and Demangeat C
1998 {\it Comput. Mater. Sci.} {\bf 10} 265

\bibitem{eriksson1991}
Eriksson O, Albers R C and A. M. Boring A M
1991 {\it Phys. Rev. Lett.} {\bf 66} 1350

\bibitem{blugel1992}
Bl\"ugel S
1992 {\it Europhys. Lett.} {\bf 18} 257

\bibitem{garcia1999}
Garc\'{\i}a A E, Gonz\'alez-Robles V and Baquero R
1999 {\it Phys. Rev. }B {\bf 59} 9392 

\bibitem{beckmann1997}
Beckmann H and Bergmann G
1997 {\it Phys. Rev. }B {\bf 55} 14350 

\bibitem{chado2001}
Chado I, Scheurer F and Bucher J P
2001 {\it Phys. Rev. }B {\bf 64} 094410

\bibitem{goldoni2001}
Goldoni A, Baraldi A, Comelli G, Esch F, Larciprete R, Lizzit S 
and G. Paolucci G
2001 {\it Phys. Rev. }B {\bf 63} 035405 

\bibitem{zhu1991}
Zhu M J, Bylander D M and Kleinman L
1991 {\it Phys. Rev. }B {\bf 43} 4007 

\bibitem{redinger1995}
Redinger J, Bl\"ugel S and Podloucky R
1995 {\it Phys. Rev. }B {\bf 51} 13852 

\bibitem{niklasson1997}
Niklasson A M N, Mirbt S, Skriver H L and Johansson B
1997 {\it Phys. Rev. }B {\bf 56} 3276 

\bibitem{galicia1993}
Galicia R 
1993 R. Mex. Fis. {\bf 32} 51 

\bibitem{reddy1993}
Reddy B V, Khanna S N and Dunlap B I
1993 {\it Phys. Rev. Lett.} {\bf 70} 3323   

\bibitem{vitos2000}
Vitos L, Johansson B and Kollar J
2000 {\it Phys. Rev. }B {\bf 62} 11957 

\bibitem{moseler2001}
Moseler M,  H\"akkinen H, Barnett R N and Landman U 
2001 {\it Phys. Rev. Lett.} {\bf 86} 2545 

\bibitem{cox1994}
Cox A J, Louderback J G, Apsel S E and Bloomfield L A
1994 {\it Phys. Rev. }B {\bf 49} 12295

\bibitem{sampedro2003}
Sampedro B, Crespo P, Hernando A, Litr\'an R,  S\'anchez L\'opez J C, 
L\'opez Cartes C, Fernandez A, Ram\'{\i}rez J, Gonz\'alez Calbet J 
and Vallet M
2003 {\it Phys. Rev. Lett.} {\bf 91} 237203 

\bibitem{taniyama1997}
Taniyama T, Ohta E and Sato T
1997 {\it Europhys. Lett.} {\bf 38} 195 

\bibitem{bazhanov2000}
Bazhanov D I, Hergert W, Stepanyuk V S, Katsnelson A A,
Rennert P, Kokko K and Demangeat C
2000 {\it Phys. Rev. }B {\bf 62} 6415

\bibitem{bellini2001}
Bellini V, Papanikolaou N, Zeller R  and Dederichs P H
2001 {\it Phys. Rev. }B {\bf 64} 094403 

\bibitem{spisak2003}
Spi{\v s}\'ak  D and J. Hafner J
2003 {\it Comp. Mat. Sci.} {\bf 27} 138;
Spi{\v s}\'ak D and J. Hafner J
2003 {\it Phys. Rev. B} 67 214416

\bibitem{zabala1998}
Zabala N, Puska M J and Nieminen R M
1998 {\it Phys. Rev. Lett.} {\bf 80} 3336; 
1999 Comment and Reply, {\it ibid.} {\bf 82} 3000 
	 
\bibitem{delin_5d_wires}
Delin A and Tosatti E
2003 {\it Phys. Rev. }B {\bf 68} 144434;
Delin A and Tosatti E 
2004 {\it Surf. Sci.} {\bf 566-568} 262

\bibitem{delin_pd_wire}
Delin A, Tosatti E and Weht R
2004 {\it Phys. Rev. Lett.} {\bf 92} 057201

\bibitem{gambardella2002}
Gambardella P, Dallmeyer A, Maiti K, Malagoli M C, Eberhardt W, Kern K and Carbone C
2002 {\it Nature (London)} {\bf 416} 301

\bibitem{dft}
Hohenberg P and Kohn W 
1964 {\it Phys. Rev.} {\bf 136} B864;
Kohn W and Sham L J 
1965 {\it Phys. Rev.} {\bf 140} A1133

\bibitem{wills}
Wills J M, Eriksson O, Alouani M and Price O L 
2000 in {\it Electronic Structure and Physical Properties of Solids},
ed Dreyss\'e  H (Berlin: Springer)

\bibitem{gga}
Perdew J P, Burke K and Ernzerhof M
1996 {\it Phys. Rev. Lett.} {\bf 77} 3865;
Perdew J P, Burke K and Ernzerhof M
1997 {\it Phys. Rev. Lett.} {\bf 78} 1396;
Zhang  Y and Yang W,
1998 {\it Phys. Rev. Lett.} {\bf 80} 890;
Perdew J P, Burke K and Ernzerhof M
1998 {\it Phys. Rev. Lett.} {\bf 80} 891

\bibitem{lda}
Ceperley D M and Alder B J
1980 {\it Phys. Rev. Lett.} {\bf 45} 566;
Perdew J P and Zunger A
1981 {\it Phys. Rev. B} {\bf 23} 5048

\bibitem{lapw}
Singh D J 1994 {\it Planewaves, pseudopotentials, and the LAPW method}
(Boston: Kluwer Academic)

\bibitem{wien97}
Blaha P, Schwarz K and Luitz J 
1997 {\it Computer code WIEN97 (Vienna University of Technology, Vienna)} 
[Improved and updated UNIX version of the original copyrighted WIEN code, 
which was published by
Blaha P, Schwarz K, Sorantin P and Trickey S B 
1990 {\it Comput. Phys. Commun.}  {\bf 59} 339]

\bibitem{xcrysden}
This figure was produced using the XCrySDen package 
available from {\tt http://www.xcrysden.org/}.
Kokalj T 1999 {\it J. Mol. Graphics Modelling} {\bf 17} 176 

\bibitem{bahn2001}
Bahn S R and Jacobsen K W
2001 {\it Phys. Rev. Lett.} {\bf 87} 266101

\bibitem{ayuela2002}
Ayuela A, Raebiger H, Puska M J and Nieminen R M
2002 {\it Phys. Rev. }B {\bf 66} 035417

\bibitem{enomoto2002}
Enomoto A, Kurokawa S and Sakai A
2002 {\it Phys. Rev. }B {\bf 65} 125410

\bibitem{ono1999}
Ono T, Ooka Y, Miyajima H and Otani Y
1999 {\it Appl. Phys. Lett.} {\bf 75} 1622

\bibitem{costi}
see, e.g., Costi T A
2000 {\it Phys. Rev. Lett.} {\bf 85} 1504

\bibitem{untiedt2003}
Untiedt C, Dekker D M T, Djukic D and van Ruitenbeek J M
2004 {\it Phys. Rev. }B {\bf 69} 081401

\bibitem{smogunov2002}
Smogunov A, Dal Corso A and Tosatti E
2002 {\it Surf. Science} {\bf 507} 609

\end{thebibliography}
 \end{document}